# Cui Prodest? Reciprocity of collaboration measured by Russian Index of Science Citation

Vladimir Pislyakov[1], Olga Moskaleva[2] and Mark Akoev[3]

[1] *pislyakov@hse.ru*
Library, National Research University Higher School of Economics, Myasnitskaya 20, Moscow, 101000 (Russia)

[2] *o.moskaleva@spbu.ru*
Saint Petersburg State University, Universitetskaya emb. 7-9, St. Petersburg, 199034 (Russia)

[3] *m.a.akoev@urfu.ru*
Ural Federal University, Mira str. 19, Ekaterinburg, 620002 (Russia)

**Abstract**
Scientific collaboration is often not perfectly reciprocal. Scientifically strong countries/institutions/laboratories may help their less prominent partners with leading scholars, or finance, or other resources. What is interesting in such type of collaboration is that (1) it may be measured by bibliometrics and (2) it may shed more light on the scholarly level of both collaborating organizations themselves. In this sense measuring institutions in collaboration sometimes may tell more than attempts to assess them as stand-alone organizations. Evaluation of collaborative patterns was explained in detail, for example, by Glänzel (2001; 2003). Here we combine these methods with a new one, made available by separating 'the best' journals from 'others' on the same platform of Russian Index of Science Citation (RISC). Such sub-universes of journals from 'different leagues' provide additional methods to study how collaboration influences the quality of papers published by organizations.

**Introduction**

This paper is a next stage in deep analysis of a relatively new national citation database RISC (Russian Index of Science Citation, launched in 2005). In the article by Moskaleva et al. (2018) the main characteristics and internal structure of RISC were for the first time presented to international audience. Next, our conference paper (Akoev, Moskaleva & Pislyakov, 2018) discussed the essential findings about publication profiles of Russian organizations obtained from the RISC database. Rather intriguing facts were revealed, but it was an analysis of universities/research institutes as stand-alone units.

Here we explore collaboration between higher education institutions, their interdependence and their roles in this interdependence.

We briefly summarize here for a reader the results of our previous works which are essential for this paper and present new findings on collaborative performance of universities.

What is essential and motivated us in this series of studies:

(1) the 'segregation by databases' method sometimes is more tangible than citation analysis. The latter, of course, has theoretically greater potential but it needs longer time lag and more citation database accuracy to be reliable;

(2) these methods are not focused on a single nation case (Russia) but may be used for world literature as soon as 'publication sub-universes' are already established by such sub-databases as other national indexes on the Web of Science (WoS) platform or even WoS Emerging Sources Citation Index as opposed to SCIE+SSCI+AHCI, 'main journal indexes'. One example of this international technique will be presented later in our paper.

*RISC and RISC Core*

Russian Index of Science Citation (RISC) was launched in 2005 on the platform eLIBRARY.RU of the Russian company *Scientific Electronic Library*. As of January 2019, RISC contains more than 31 mln documents with more than 360 mln references, the total list of journals contains more than 60,000 titles, about 6,000 of them are fully indexed in RISC, other journals are sources of documents with Russian affiliations and papers citing them. Along with journals, RISC indexes conference proceedings, books, patents, dissertations and other research artefacts.

As it follows from its name, RISC main feature is citation indexing and analytics. However, in our study of collaborations we will not use citation analysis. What is more important for the present paper is that in 2017 a special subset of journals was defined, RISC Core. It contains those RISC journals which are also indexed either in Web of Science or in Scopus. It is a kind of 'best journals' subset of RISC periodicals.

It should be mentioned that since 2016 there is a special national citation index on the Web of Science platform, Russian Science Citation Index with about 800 journals (RSCI, not to be confused with RISC which is a much bigger database on the separate eLIBRARY.RU platform, with no connection to WoS). Journals for RSCI were selected through professional expertise by specially summoned Russian academia board and by expert opinion of top Russian scientists, more than 12,500 of them voted for journals. All RSCI journals are also included into RISC Core subset. It is important as soon as makes RISC Core independent from decisions made by foreign commercial companies Clarivate Analytics and Elsevier on inclusion/non-inclusion of Russian titles into their databases. This adds value to RISC Core as an instrument for analysis of Russian organizations' performance.

For assessment of the RISC Core a brief experiment was made. It was found that RISC Core which contained in 2016 only 23% of papers of all eLIBRARY.RU platform has attracted 93% of all citations there. This affirms the quality of the RISC Core sources and their status of, so to say, 'best journals'.

Further details on RISC may be found in (Moskaleva et al., 2018; Akoev, Moskaleva & Pislyakov, 2018).

**Data and overview of previous results**
We study 16 Russian universities and their output indicators taken from RISC database. These universities were chosen by three criteria:

(1) they should be leaders by number of documents in RISC as a whole;
(2) they should be leaders by number of documents in RISC Core;
(3) they should be involved into collaboration with the most of other chosen universities.

In tables and graphs we use abbreviated names of these institutions, but in the text we will usually spell them out. The full list of the universities with their abbreviations may be found in Appendix 1. We omit those collaborations which resulted in less than 20 papers in the whole RISC and regard these data as insufficient in all tables.

Output indicators of these organizations for the whole RISC and for the RISC Core are taken for five-year 2013–2017 period from the predefined reports in their profiles at eLIBRARY.RU.

The same approach was used by Akoev, Moskaleva and Pislyakov (2018) for 2012–2016 time period. Here we reproduce a graph from there (Figure 4 there, Figure 1 here), most important for the current analysis. The graph is corrected so that only universities studied in this paper are shown.

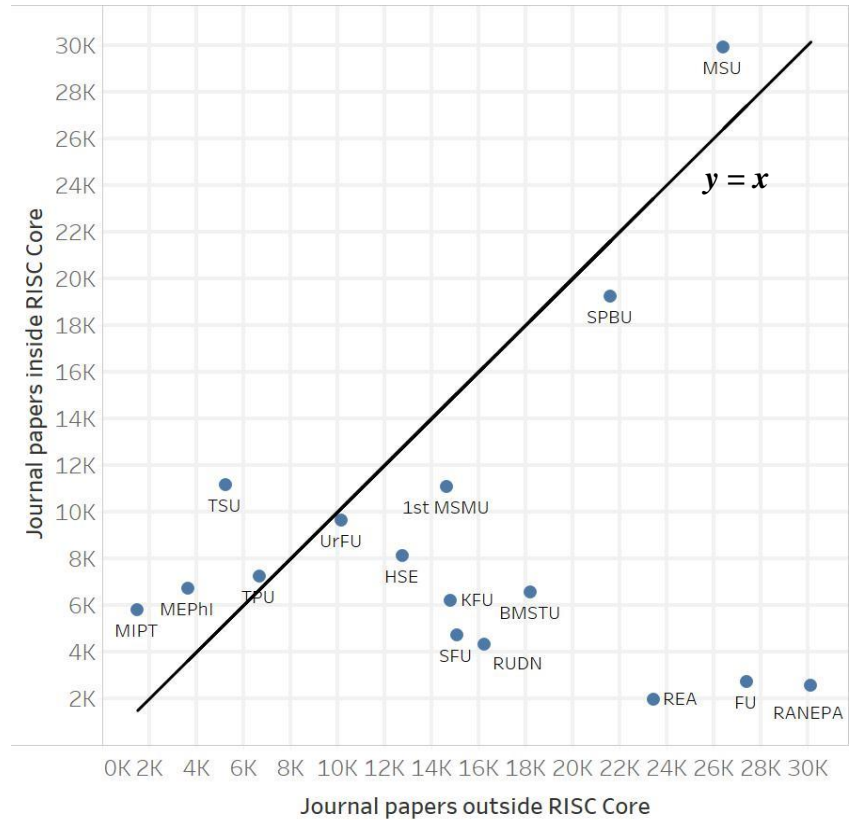

**Figure 1. Most prolific Russian Universities: journal papers inside and outside RISC Core. (adapted from Akoev, Moskaleva and Pislyakov, 2018)**

We can see that for different universities the share of documents in the Core is drastically different. The highest percentage of the Core papers is found for Moscow Institute of Physics and Technology (MIPT), Moscow Engineering Physics Institute (MEPhI) and Tomsk State University (TSU). The lowest share of their papers in 'the best journals' have Academy of National Economy (RANEPA), Plekhanov University of Economics (REA) and Financial University (FU). This previous analysis of the universities as stand-alone organizations will be useful in our study on collaborations.

**Results and discussion**

*Symmetrical (bi-directional) collaboration indexes*

First, we study simple symmetrical indexes measuring the strength of collaboration between Russian universities. Generally, two such indexes may be used, Jaccard index or Salton measure (Glänzel, 2003). We use here Jaccard and calculate it both for collaboration in the whole RISC and only in RISC Core.

Jaccard index is a ratio of the cardinality of the intersection of two sets to the cardinality of the union of these sets, expressed as a percentage. In our case, it is the number of papers written by both organizations jointly divided by the number of papers written at least by one of these organizations. This index gives an estimate of the relative strength of collaboration between two organizations, normalized in a way by the volume of their outputs. Five joint papers are a weaker level of the collaboration when each of the institutions has 1000 articles than when they have published only 10 papers each. Jaccard aims to correct for this difference.

To illustrate, the highest Jaccard index in our data is found for collaboration of two Tomsk universities (State and Polytechnic, TSU and TPU) in the RISC Core — 7.22%. This means that every 14th paper (100/7.22) written by scientists *either* from TSU *or* TPU in the RISC Core journals is *jointly* written by them. The full data on Jaccard indices may be found in Appendix 2.

For our study, the Jaccard indices themselves are not so interesting as their ratio for collaboration in the RISC Core and RISC as a whole. It will reveal the change of the importance of collaboration between two organizations when moving from the entire database to 'the best journals'. Table 1 shows Jaccards for inside-core collaboration of each pair of universities divided by Jaccard for their co-authorship in the whole RISC. As soon as Jaccard is a symmetrical indicator, the matrix in Table 1 is also symmetrical.

**Table 1. Ratio of Jaccard indices for collaboration in RISC Core/whole RISC.**
**(light blue cell: <1; bold font: >2; yellow cell: >3; empty cell: insufficient data)**

|          | MSU  | FU   | SPBU | RANEPA | REA  | UrFU | SFU  | RUDN | BMSTU | KFU  | TSU  | 1st MSMU | HSE  | TPU  | MEPhI | MIPT |
|----------|------|------|------|--------|------|------|------|------|-------|------|------|----------|------|------|-------|------|
| MSU      |      | 0.81 | 1.80 | 1.13   | 0.82 | 2.06 | 1.54 | 1.12 | 1.87  | 1.97 | 2.14 | 1.61     | 1.49 | 2.09 | 2.33  | 2.16 |
| FU       | 0.81 |      | 1.24 | 1.04   | 1.97 | 0.66 | 1.62 | 1.08 | 1.94  | 2.66 |      | 4.65     | 1.91 |      | 2.43  | 2.71 |
| SPBU     | 1.80 | 1.24 |      | 1.02   | 1.01 | 1.85 | 1.77 | 1.32 | 2.41  | 2.32 | 2.08 | 1.28     | 1.43 | 1.26 | 2.47  | 2.42 |
| RANEPA   | 1.13 | 1.04 | 1.02 |        | 1.31 | 0.55 | 0.64 | 2.27 | 0.98  | 0.82 | 0.23 | 1.38     | 2.55 |      | 1.24  |      |
| REA      | 0.82 | 1.97 | 1.01 | 1.31   |      | 0.68 | 1.26 | 2.82 | 1.67  | 3.23 |      | 1.28     | 0.99 |      | 3.38  |      |
| UrFU     | 2.06 | 0.66 | 1.85 | 0.55   | 0.68 |      | 2.34 |      |       | 2.26 | 1.46 |          | 1.26 | 2.08 | 1.79  | 2.10 |
| SFU      | 1.54 | 1.62 | 1.77 | 0.64   | 1.26 | 2.34 |      | 1.70 |       | 2.17 | 1.54 |          | 1.10 |      | 2.26  | 2.53 |
| RUDN     | 1.12 | 1.08 | 1.32 | 2.27   | 2.82 |      |      |      | 1.26  | 3.04 | 1.65 | 1.35     | 0.62 |      | 2.43  | 2.66 |
| BMSTU    | 1.87 | 1.94 | 2.41 | 0.98   | 1.67 |      | 1.70 | 1.26 |       | 0.52 | 2.01 | 1.84     | 1.64 | 2.25 | 1.83  | 1.89 |
| KFU      | 1.97 | 2.66 | 2.32 | 0.82   | 3.23 | 2.26 | 2.17 | 3.04 | 0.52  |      |      | 2.35     | 0.87 |      | 2.32  | 2.53 |
| TSU      | 2.14 |      | 2.08 | 0.23   |      | 1.46 | 1.54 | 1.65 | 2.01  |      |      |          | 1.16 | 1.83 | 2.03  | 1.81 |
| 1st MSMU | 1.61 | 4.65 | 1.28 | 1.38   | 1.28 |      |      | 1.35 | 1.84  | 2.35 |      |          | 1.71 |      |       | 1.40 |
| HSE      | 1.49 | 1.91 | 1.43 | 2.55   | 0.99 | 1.26 | 1.10 | 0.62 | 1.64  | 0.87 | 1.16 | 1.71     |      |      | 1.74  | 1.93 |
| TPU      | 2.09 |      | 1.26 |        |      | 2.08 |      | 2.25 |       |      | 1.83 |          |      |      | 1.68  |      |
| MEPhI    | 2.33 | 2.43 | 2.47 | 1.24   | 3.38 | 1.79 | 2.26 | 2.43 | 1.83  | 2.32 | 2.03 |          | 1.74 | 1.68 |       | 1.77 |
| MIPT     | 2.16 | 2.71 | 2.42 |        |      | 2.10 | 2.53 | 2.54 | 1.89  | 2.53 | 1.81 | 1.40     | 1.93 |      | 1.77  |      |

There are only 13 out of 120 collaborative pairs in Table 1 for which ratio of Jaccard indices is less than one (these are shown as light blue cells). This means that for 89% pairs of universities their collaboration inside RISC Core is more important than in all other journals in RISC. Extramural collaboration plays a greater role in high-level research. A similar behaviour of Jaccard indices was found by Pislyakov and Shukshina (2014) for international collaboration of Russian organizations in producing highly cited papers.

Most remarkable are four pairs shown in Table 1 as yellow cells. These are Plekhanov University of Economics (REA) in collaboration with Kazan Federal University (KFU) and Moscow Engineering Physics Institute (MEPhI); Peoples' Friendship University (RUDN) again with Kazan Federal University (KFU); and the highest value in Table 1 is for the pair Financial University (FU)/Moscow Medical University (1st MSMU). In the latter case collaboration of these institutions in the core is 4.65 times more important for them than in the whole RISC. They do not collaborate actively, but when they do so, most of their joint papers become published in the core (15 out of 20).

There are five organizations which become more important for any partner when the observation moves to the core publications. These are St. Petersburg University, Moscow Medical University, Tomsk Polytechnic University, Moscow Engineering Physics Institute and Moscow Institute of Physics and Technology. The latter, MIPT, also has the highest *average* ratio of Jaccards in Table 1 (2.16). And conversely (as we speak of bi-directional indicator), for these universities each other collaborator is also more important in the RISC Core than in the RISC as a whole database.

On the other side one can notice Academy of National Economy (RANEPA) performance. There are five cases (out of 13) of weakened collaboration when moving to the Core. Collaboration is not so important for RANEPA when preparing papers for the leading journals. Or, maybe, this relative negligence of collaboration in the RISC Core is a cause of the most unbalanced position of RANEPA concerning inside/outside Core publications observed in Figure 1.

*Asymmetrical indexes: collaborative gain*

More profound analysis of the patterns of inter-university collaboration may be obtained by investigating asymmetrical, unidirectional indexes which show how profitable is the collaboration for one partner and for another. Here we use an indicator of 'collaborative gain'. It is a ratio between percentage of papers found in the RISC Core for joint papers written by two institutions and share of Core papers in total output by one of them. In calculating the collaborative gain the numerator of the index is the same for both partner universities, it is a share of their collaborative papers which have entered the Core. But the denominator is different as soon as it is a share of all Core papers for a given institution. That is why A-B index is different from B-A.

**Table 2. Collaborative gain received by university in a row from university in a column.**
**(light blue cell: <1; empty cell: insufficient data)**

| | MSU | FU | SPBU | RANEPA | REA | UrFU | SFU | RUDN | BMSTU | KFU | TSU | 1st MSMU | HSE | TPU | MEPhI | MIPT |
|---|---|---|---|---|---|---|---|---|---|---|---|---|---|---|---|---|
| MSU | | 0.48 | 1.72 | 0.69 | 0.55 | 1.97 | 1.25 | 0.90 | 1.59 | 1.77 | 2.29 | 1.64 | 1.44 | 2.00 | 2.36 | 2.36 |
| FU | 3.82 | | 5.02 | 1.05 | 2.06 | 2.10 | 3.03 | 2.00 | 4.07 | 6.64 | | 15.71 | 5.61 | | 7.27 | 8.20 |
| SPBU | 1.90 | 0.70 | | 0.60 | 0.65 | 1.83 | 1.43 | 1.05 | 2.06 | 2.12 | 2.34 | 1.37 | 1.44 | 1.24 | 2.63 | 2.80 |
| RANEPA | 5.38 | 1.03 | 4.19 | | 1.38 | 1.78 | 1.21 | 4.23 | 2.08 | 2.09 | 0.89 | 4.77 | 7.60 | | 3.79 | |
| REA | 3.82 | 1.80 | 4.05 | 1.23 | | 2.19 | 2.35 | 5.17 | 3.53 | 8.18 | | 4.47 | 3.02 | | 10.64 | |
| UrFU | 2.25 | 0.30 | 1.89 | 0.26 | 0.36 | | 1.74 | | | 2.03 | 1.75 | | 1.28 | 2.05 | 1.99 | 2.63 |
| SFU | 2.92 | 0.89 | 3.02 | 0.36 | 0.79 | 3.56 | | | 1.89 | 2.78 | 2.85 | | 1.66 | | 3.63 | 4.49 |
| RUDN | 2.17 | 0.61 | 2.28 | 1.31 | 1.80 | | | | 1.42 | 3.95 | 3.10 | 2.29 | 0.94 | | 3.95 | 4.73 |
| BMSTU | 2.99 | 0.96 | 3.50 | 0.50 | 0.96 | | 1.52 | 1.11 | | 0.59 | 3.30 | 2.74 | 2.19 | 2.89 | 2.64 | 3.06 |
| KFU | 2.59 | 1.22 | 2.80 | 0.39 | 1.73 | 2.59 | 1.74 | 2.39 | 0.46 | | | 2.99 | 1.00 | | 2.93 | 3.61 |
| TSU | 1.80 | | 1.66 | 0.09 | | 1.20 | 0.96 | 1.01 | 1.38 | | | | 0.99 | 1.47 | 1.93 | 2.02 |
| 1st MSMU | 1.51 | 1.82 | 1.13 | 0.56 | 0.59 | | | 0.87 | 1.34 | 1.88 | | | 1.57 | | | 1.67 |
| HSE | 1.58 | 0.77 | 1.42 | 1.07 | 0.48 | 1.23 | 0.78 | 0.42 | 1.27 | 0.75 | 1.38 | 1.87 | | | 1.92 | 2.46 |
| TPU | 2.35 | | 1.31 | | | 2.11 | | | 1.80 | | 2.20 | | | | 1.91 | |
| MEPhI | 2.07 | 0.80 | 2.08 | 0.43 | 1.35 | 1.52 | 1.36 | 1.43 | 1.23 | 1.76 | 2.15 | | 1.54 | 1.43 | | 2.09 |
| MIPT | 1.30 | 0.56 | 1.38 | | | 1.26 | 1.05 | 1.02 | 0.89 | 1.35 | 1.41 | 0.99 | 1.23 | | 1.31 | |
| average | 2.56 | 0.92 | 2.50 | 0.66 | 1.06 | 1.94 | 1.53 | 1.80 | 1.79 | 2.76 | 2.15 | 3.88 | 2.25 | 1.85 | 3.49 | 3.34 |

Collaborative gain indicator for 16 universities we study are summarized in Table 2. Collaborative gain may easily be interpreted in such a way: how more often university A publishes its research in 'the best journals' if it collaborates with the university B? For example, value 3.82 in the row 3, column 2 cell of Table 2 means that collaborative papers by Financial University (FU) and Moscow University (MSU) have a 3.82 times higher share in RISC Core than all papers by FU.

Columns in Table 2 represent the 'gain', 'profit' which an organization brings to its partner. Cells marked by light blue color contain values less than one, this means that the gain is negative. In a sense this means that collaboration with this university (in a column) hinders its partner university (in a row). The probability of publishing such collaborative research in high-quality journals is lower than average for organization in a row.

We may notice that six out of 16 universities are *always* positive partners (see their columns in Table 2). Each other organization receives profit from making collaborative research with them. This is, of course, a natural situation as soon as collaboration should bring success. However, several institutions often bring difficulties with their partnership. These 'negative leaders' are Financial University (FU) and Academy of National Economy (RANEPA) with nine 'negative' cells in their columns, Plekhanov University of Economics (REA) following them with seven negative cases. Note that all three universities are among weakest in Figure 1. And they all are focused more on social sciences. This phenomenon should be scrutinized in further research.

The last row of Table 2 contains averages for all indexes for a given university in a column. Although these averages are somewhat artificial, still they may provide some insight into the overall value of the organization as a partner. The data from this row of Table 2 are sorted and organized into Figure 2.

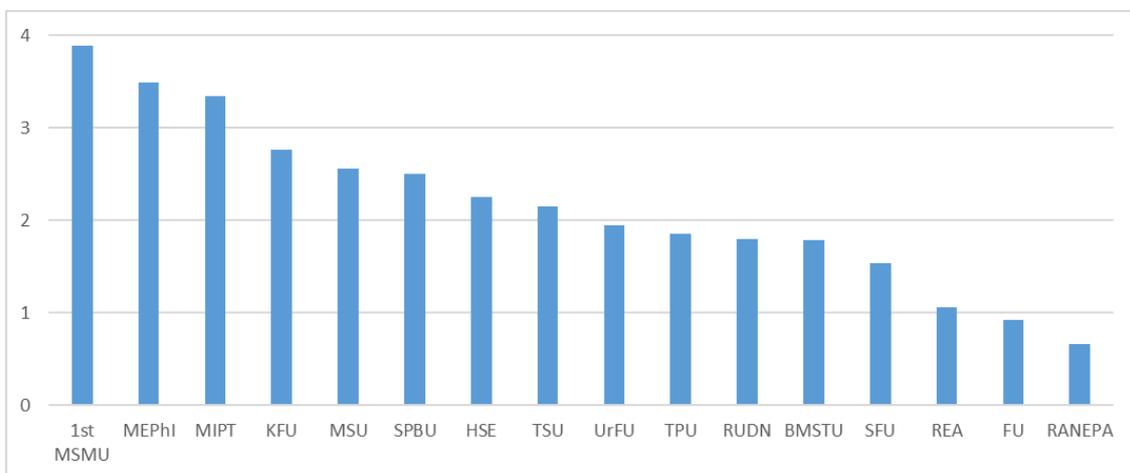

**Figure 2. Average collaborative gain a university brings to its partners.**

Finally, Table 2 also shows that there are three organizations which collaborations always bring gain *to themselves*. They are Financial University (FU), Plekhanov University of Economics (REA) and Tomsk Polytechnical University (TPU), their rows do not contain blue cells. The profitable strategy for these institutions is further strengthening of extramural collaborations.

*Reciprocity of collaboration*

Generally, collaboration is a win-win process when each of the partners receives some surplus compared to its sole performance. We observe such situation for the majority of collaborative pairs in Table 2 when both A-B and B-A cells contain values greater than one and in this sense their partnership is *reciprocal*, profitable for both sides.

However, there are some remarkable exceptions. There are three pairs when collaboration is not very successful for both sides: Bauman University (BMSTU) with Kazan University (KFU); Tomsk State University (TSU) with Academy of National Economy (RANEPA); and the Higher School of Economics (HSE) with Peoples' Friendship University (RUDN). In each case there is negative collaborative effect for both institutions.

More often another situation may be found—when one of the partners gains from collaboration while another side loses. There are 9 cases of such kind of 'collaborative vampirism' for Financial University (FU), 8 for Academy of National Economy (RANEPA), 7 for Plekhanov University of Economics (REA), 2 for both Southern Federal University (SFU) and Peoples' Friendship University (RUDN), one for Bauman University (BMSTU), Kazan Federal University (KFU), Moscow Medical University (1st MSMU) and the Higher School of Economics (HSE). These are the cases of *non-reciprocal* inter-university collaboration, win-loss situation, when partnership benefits one side while harms another.

Still, even in win-win situations the gain one university brings to another is generally not equal to the gain it receives in return. To assess this asymmetry, we created Table 3 where differences of collaborative gains received and brought are calculated for each partnership. Positive values in Table 3 mean that university in a column has brought more gain to university in a row than it received in exchange. By definition, this table is skew-symmetric.

**Table 3. Non-reciprocity (difference in collaborative gain).**
**How university in a column is *non-reciprocal* to university in a row.**
**(blue cell: < –3; light blue cell: <–0.1; yellow cell: between –0.1 and 0.1; empty cell: insuff. data)**

|  | MSU | FU | SPBU | RANEPA | REA | UrFU | SFU | RUDN | BMSTU | KFU | TSU | 1st MSMU | HSE | TPU | MEPhI | MIPT |
|---|---|---|---|---|---|---|---|---|---|---|---|---|---|---|---|---|
| MSU |  | -3.34 | -0.19 | -4.69 | -3.27 | -0.28 | -1.67 | -1.27 | -1.40 | -0.82 | 0.49 | 0.13 | -0.14 | -0.35 | 0.29 | 1.07 |
| FU | 3.34 |  | 4.32 | 0.02 | 0.27 | 1.79 | 2.14 | 1.40 | 3.11 | 5.42 |  | 13.90 | 4.84 |  | 6.47 | 7.64 |
| SPBU | 0.19 | -4.32 |  | -3.60 | -3.40 | -0.06 | -1.59 | -1.24 | -1.44 | -0.68 | 0.68 | 0.23 | 0.02 | -0.07 | 0.55 | 1.42 |
| RANEPA | 4.69 | -0.02 | 3.60 |  | 0.15 | 1.52 | 0.84 | 2.91 | 1.58 | 1.70 | 0.80 | 4.20 | 6.53 |  | 3.37 |  |
| REA | 3.27 | -0.27 | 3.40 | -0.15 |  | 1.83 | 1.56 | 3.37 | 2.57 | 6.46 |  | 3.87 | 2.54 |  | 9.29 |  |
| UrFU | 0.28 | -1.79 | 0.06 | -1.52 | -1.83 |  | -1.82 |  |  | -0.56 | 0.55 |  | 0.05 | -0.05 | 0.46 | 1.37 |
| SFU | 1.67 | -2.14 | 1.59 | -0.84 | -1.56 | 1.82 |  |  | 0.37 | 1.04 | 1.89 |  | 0.88 |  | 2.27 | 3.44 |
| RUDN | 1.27 | -1.40 | 1.24 | -2.91 | -3.37 |  |  |  | 0.31 | 1.56 | 2.09 | 1.42 | 0.51 |  | 2.52 | 3.71 |
| BMSTU | 1.40 | -3.11 | 1.44 | -1.58 | -2.57 |  | -0.37 | -0.31 |  | 0.13 | 1.92 | 1.40 | 0.92 | 1.09 | 1.41 | 2.17 |
| KFU | 0.82 | -5.42 | 0.68 | -1.70 | -6.46 | 0.56 | -1.04 | -1.56 | -0.13 |  |  | 1.11 | 0.25 |  | 1.17 | 2.26 |
| TSU | -0.49 |  | -0.68 | -0.80 |  | -0.55 | -1.89 | -2.09 | -1.92 |  |  |  | -0.39 | -0.73 | -0.22 | 0.61 |
| 1st MSMU | -0.13 | -13.90 | -0.23 | -4.20 | -3.87 |  |  | -1.42 | -1.40 | -1.11 |  |  | -0.30 |  |  | 0.68 |
| HSE | 0.14 | -4.84 | -0.02 | -6.53 | -2.54 | -0.05 | -0.88 | -0.51 | -0.92 | -0.25 | 0.39 | 0.30 |  |  | 0.38 | 1.23 |
| TPU | 0.35 |  | 0.07 |  |  | 0.05 |  |  | -1.09 |  | 0.73 |  |  |  | 0.48 |  |
| MEPhI | -0.29 | -6.47 | -0.55 | -3.37 | -9.29 | -0.46 | -2.27 | -2.52 | -1.41 | -1.17 | 0.22 |  | -0.38 | -0.48 |  | 0.79 |
| MIPT | -1.07 | -7.64 | -1.42 |  |  | -1.37 | -3.44 | -3.71 | -2.17 | -2.26 | -0.61 | -0.68 | -1.23 |  | -0.79 |  |
| average | 1.03 | -4.20 | 0.89 | -2.45 | -3.14 | 0.40 | -0.87 | -0.58 | -0.28 | 0.73 | 0.83 | 2.59 | 1.01 | -0.10 | 1.97 | 2.20 |

There are several pairs with almost perfect reciprocity, for example FU/RANEPA. Next, SPBU, UrFU, HSE and TPU also collaborate almost perfectly reciprocally with each other. We should remind, however, that according to our definitions the reciprocal collaboration may exist only between organizations of equal strength, in terms of share of their papers in the RISC Core (when denominators in the formula for collaborative gain are equal).

For each partner MIPT brings more gain than receives from them. The opposite is the case of Financial University (FU). Again, overall relative non-reciprocity of a given institution may be estimated with the column average of Table 3. Results are shown in Figure 3. This graph does not necessarily follow the sorting order from the higher to lower share of university's documents in RISC (as may be suggested by mathematical definition of 'gain'), at least because there are empty cells in Table 3, not all collaborations are active.

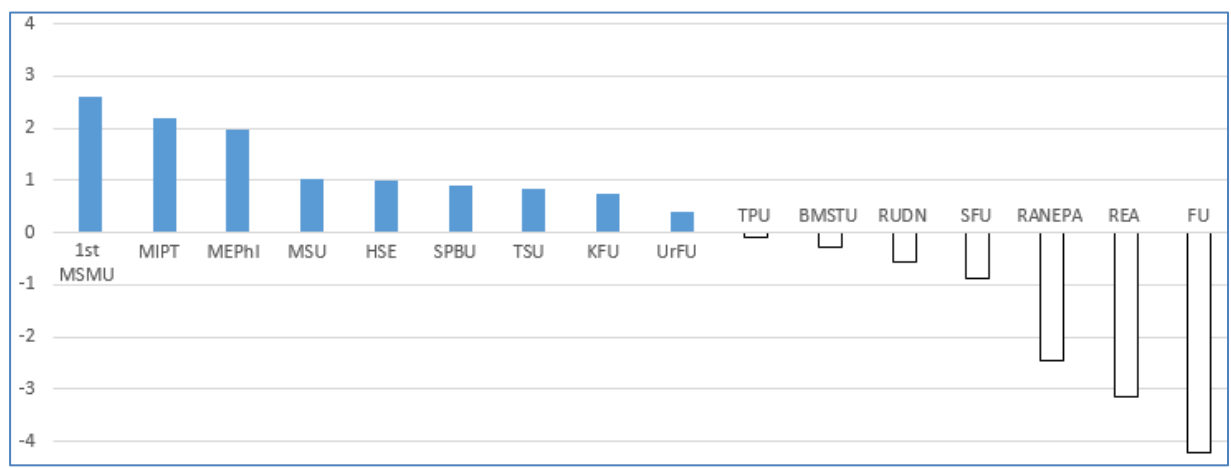

**Figure 3. Average non-reciprocity in collaboration (how much a university brings to collaborations compared to how much it receives).**

One may say that here we have nine 'donors', six 'acceptors' and one university (Tomsk Polytechnical) on the borderline. Of course, it is only an approximation and, additionally, made on the limited number of partner universities. Still it may shed light on the collaborative behavior and collaborative gains of the institutions we study. For example, the most significant change is for Kazan Federal University—from 4th place in Figure 2 to 8th in Figure 3. What does it mean? It means that KFU brings much gain in collaborations, but tends to collaborate mainly with those universities, who also give in return. As a result, it receives a lot too. Perhaps this is the most balanced and prudent though not altruistic approach to scientific collaboration.

**Conclusion**

We have found that study of collaboration between universities shows them in a more detailed manner, revealing their standing and roles in a complex network of collective scientific activity. We conclude that, in general, organizations strong by themselves are also strong collaborators. On the contrary, the universities with low share of papers published in the high-level journals are not very beneficial partners to the others. However, there are some institutions which break this pattern, for example one of the best collaborators appeared to be Moscow Medical University (1st MSMU), which has quite moderate position in Figure 1 showing performance of universities per se.

Our method of quality analysis by distribution of papers across journal subsets of different scientific level may be generalized. For example, those studies which analyze scattering of publications across journal quartiles use the same principle. They may focus only on the highest (first) quartile papers as in (Chowdhury & Philipson, 2016) or on full distribution by quartile (Olmeda-Gómez & Moya-Anegón, 2016). Quartiles may come from impact factor (Bordons & Barrigón, 1992), or SJR (Chinchilla-Rodríguez et al., 2015), or some other rankings. The idea of the method remains the same.

Moreover, this approach may be applied to Web of Science databases which have different status—namely Emerging Sources Citation Index as opposed to SCIE+SSCI+AHCI. As a small exercise, we have gathered statistics for different countries on how they publish their collaborative papers in different subsets of Web of Science. It appeared that, for example, collaboration of France and Germany is rather reciprocal and benefits both sides: joint Web of Science papers of these countries enter the 'main' journal indexes 2.8% and 2.6% more often than all publications of France and Germany, respectively. But at the same time when China partners with Iran, it publishes 5.1% more papers in 'emerging' index of WoS, while Iran gets 7.0% more its papers in the 'main' SCIE+SSCI+AHCI indexes, compared to its total WoS output. This is an example of non-reciprocal collaboration of countries measured with the international citation database.

As next steps of our research we plan to focus on disciplinary differences in collaboration and investigate cross-collaboration patterns between universities and Russian research institutes as observed from Russian Index of Science Citation and its Core.

**Appendix 1. List of universities' abbreviations.**

| Abbreviation | Full name of university |
|---|---|
| 1st MSMU | Sechenov First Moscow State Medical University |
| BMSTU | Bauman Moscow State Technical University |

| Abbreviation | Full name of university |
|---|---|
| FU | Financial University under the Government of the Russian Federation |
| HSE | National Research University Higher School of Economics |
| KFU | Kazan Volga Region Federal University |
| MEPhI | Moscow Engineering Physics Institute |
| MIPT | Moscow Institute of Physics and Technology |
| MSU | Moscow State University |
| RANEPA | Russian Presidential Academy of National Economy and Public Administration |
| REA | Plekhanov Russian University of Economics |
| RUDN | Peoples' Friendship University of Russia |
| SFU | Southern Federal University |
| SPBU | St. Petersburg State University |
| TPU | Tomsk Polytechnic University |
| TSU | Tomsk State University |
| UrFU | Ural Federal University |

**Appendix 2. Jaccard indices for inter-university collaboration in the entire RISC and RISC Core, % (empty cell: insufficient data).**

| | MSU | | FU | | SPBU | | RANEPA | | REA | | UrFU | | SFU | | RUDN | | BMSTU | | KFU | | TSU | | 1st MSMU | | HSE | | TPU | | MEPhI | | MIPT | |
|---|---|---|---|---|---|---|---|---|---|---|---|---|---|---|---|---|---|---|---|---|---|---|---|---|---|---|---|---|---|---|---|---|
| | RISC | Core | RISC | Core | RISC | Core | RISC | Core | RISC | Core | RISC | Core | RISC | Core | RISC | Core | RISC | Core | RISC | Core | RISC | Core | RISC | Core | RISC | Core | RISC | Core | RISC | Core | RISC | Core |
| MSU | | | 0.19 | 0.15 | 0.38 | 0.69 | 0.26 | 0.29 | 0.19 | 0.15 | 0.18 | 0.36 | 0.16 | 0.25 | 0.35 | 0.39 | 0.71 | 1.32 | 0.25 | 0.49 | 0.53 | 1.14 | 0.35 | 0.56 | 0.72 | 1.07 | 0.07 | 0.16 | 1.41 | 3.29 | 1.30 | 2.82 |
| FU | 0.19 | 0.15 | | | 0.07 | 0.09 | 0.66 | 0.68 | 1.43 | 2.81 | 0.04 | 0.03 | 0.08 | 0.13 | 0.22 | 0.23 | 0.14 | 0.27 | 0.04 | 0.11 | | | 0.02 | 0.10 | 0.24 | 0.45 | | | 0.11 | 0.27 | 0.03 | 0.08 |
| SPBU | 0.38 | 0.69 | 0.07 | 0.09 | | | 0.21 | 0.21 | 0.02 | 0.02 | 0.14 | 0.27 | 0.05 | 0.09 | 0.13 | 0.16 | 0.08 | 0.20 | 0.10 | 0.24 | 0.16 | 0.33 | 0.08 | 0.11 | 0.28 | 0.40 | 0.04 | 0.05 | 0.23 | 0.58 | 0.10 | 0.23 |
| RANEPA | 0.26 | 0.29 | 0.66 | 0.68 | 0.21 | 0.21 | | | 0.97 | 1.27 | 0.12 | 0.06 | 0.15 | 0.10 | 0.29 | 0.65 | 0.07 | 0.07 | 0.05 | 0.04 | 0.03 | 0.01 | 0.09 | 0.13 | 0.35 | 0.90 | | | 0.03 | 0.04 | | |
| REA | 0.19 | 0.15 | 1.43 | 2.81 | 0.02 | 0.02 | 0.97 | 1.27 | | | 0.03 | 0.02 | 0.07 | 0.09 | 0.22 | 0.62 | 0.19 | 0.31 | 0.09 | 0.31 | | | 0.06 | 0.08 | 0.18 | 0.18 | | | 0.07 | 0.23 | | |
| UrFU | 0.18 | 0.36 | 0.04 | 0.03 | 0.14 | 0.27 | 0.12 | 0.06 | 0.03 | 0.02 | | | 0.05 | 0.12 | | | | | 0.06 | 0.13 | 0.15 | 0.22 | | | 0.09 | 0.11 | 0.06 | 0.12 | 0.08 | 0.15 | 0.08 | 0.17 |
| SFU | 0.16 | 0.25 | 0.08 | 0.13 | 0.05 | 0.09 | 0.15 | 0.10 | 0.07 | 0.09 | 0.05 | 0.12 | | | | | 0.04 | 0.06 | 0.04 | 0.09 | 0.08 | 0.13 | | | 0.04 | 0.04 | | | 0.07 | 0.15 | 0.07 | 0.19 |
| RUDN | 0.35 | 0.39 | 0.22 | 0.23 | 0.13 | 0.16 | 0.29 | 0.65 | 0.22 | 0.62 | | | | | | | 0.24 | 0.30 | 0.04 | 0.12 | 0.06 | 0.09 | 0.78 | 1.06 | 0.14 | 0.08 | | | 0.11 | 0.26 | 0.08 | 0.22 |
| BMSTU | 0.71 | 1.32 | 0.14 | 0.27 | 0.08 | 0.20 | 0.07 | 0.07 | 0.19 | 0.31 | | | 0.04 | 0.06 | 0.24 | 0.30 | | | 0.04 | 0.02 | 0.07 | 0.14 | 0.10 | 0.19 | 0.18 | 0.29 | 0.04 | 0.08 | 0.77 | 1.42 | 0.35 | 0.66 |
| KFU | 0.25 | 0.49 | 0.04 | 0.11 | 0.10 | 0.24 | 0.05 | 0.04 | 0.09 | 0.31 | 0.06 | 0.13 | 0.04 | 0.09 | 0.04 | 0.12 | 0.04 | 0.02 | | | | | 0.08 | 0.20 | 0.07 | 0.06 | | | 0.04 | 0.09 | 0.18 | 0.46 |
| TSU | 0.53 | 1.14 | | | 0.16 | 0.33 | 0.03 | 0.01 | | | 0.15 | 0.22 | 0.08 | 0.13 | 0.06 | 0.09 | 0.07 | 0.14 | | | | | | | 0.04 | 0.05 | 3.95 | 7.22 | 0.81 | 1.65 | 0.83 | 1.49 |
| 1st MSMU | 0.35 | 0.56 | 0.02 | 0.10 | 0.08 | 0.11 | 0.09 | 0.13 | 0.06 | 0.08 | | | | | 0.78 | 1.06 | 0.10 | 0.19 | 0.08 | 0.20 | | | | | 0.03 | 0.06 | | | | | 0.14 | 0.19 |
| HSE | 0.72 | 1.07 | 0.24 | 0.45 | 0.28 | 0.40 | 0.35 | 0.90 | 0.18 | 0.18 | 0.09 | 0.11 | 0.04 | 0.04 | 0.14 | 0.08 | 0.18 | 0.29 | 0.07 | 0.06 | 0.04 | 0.05 | 0.03 | 0.06 | | | | | 0.13 | 0.22 | 0.76 | 1.46 |
| TPU | 0.07 | 0.16 | | | 0.04 | 0.05 | | | | | 0.06 | 0.12 | | | | | 0.04 | 0.08 | | | 3.95 | 7.22 | | | | | | | 0.28 | 0.47 | | |
| MEPhI | 1.41 | 3.29 | 0.11 | 0.27 | 0.23 | 0.58 | 0.03 | 0.04 | 0.07 | 0.23 | 0.08 | 0.15 | 0.07 | 0.15 | 0.11 | 0.26 | 0.77 | 1.42 | 0.04 | 0.09 | 0.81 | 1.65 | | | 0.13 | 0.22 | 0.28 | 0.47 | | | 2.92 | 5.16 |
| MIPT | 1.30 | 2.82 | 0.03 | 0.08 | 0.10 | 0.23 | | | | | 0.08 | 0.17 | 0.07 | 0.19 | 0.09 | 0.22 | 0.35 | 0.66 | 0.18 | 0.46 | 0.83 | 1.49 | 0.14 | 0.19 | 0.76 | 1.46 | | | 2.92 | 5.16 | | |